\documentclass[aps,twocolumn,prx,amsfonts,amssymb,showpacs,superscriptaddress,floatfix, longbibliography]{revtex4-2}

\usepackage[T1]{fontenc}
\usepackage[utf8]{inputenc}
\usepackage{amsmath,amssymb}
\usepackage{graphicx}
\usepackage{xcolor}
\usepackage{hyperref}

\hypersetup{colorlinks=true,linkcolor=blue,citecolor=blue,urlcolor=blue}

\begin{document}

\title{Dynamical confinement creates quasi-critical scale-free dynamics}

\author{Kilian Duplat}
\author{Adèle Douin}
\author{François Detcheverry}
\author{Osvanny Ramos}
\email{osvanny.ramos@univ-lyon1.fr}
\affiliation{Universit\'e Lyon 1, CNRS, Institut Lumi\`ere Mati\`ere, UMR5306, F-69622 Villeurbanne, France}

\begin{abstract}
Scale-free avalanches are commonly interpreted as signatures of criticality. Conventional theoretical frameworks predict robust avalanche-size exponents $\tau$ bounded by $3/2$, while dissipation is expected to drive systems away from criticality, by progressively suppressing large events. Yet earthquakes and numerous dissipative experiments exhibit scale-free dynamics with non-universal exponents $\tau > 3/2$, challenging both expectations. Using the Olami–Feder–Christensen model as a minimal theoretical framework, we identify two fundamentally distinct routes to scale-free dynamics:a genuinely critical scenario, in which avalanches continuously explore the entire system and remain permanently capable of spanning it; and a quasi-critical scenario, in which dissipation and the traces left by past avalanches generate dynamical confinement. Rather than remaining at a critical point, the system explores a power-law distribution of distances to criticality. We show, using a simple analytical model, how this mechanism produces scale-free avalanche statistics with dissipation-dependent exponents exceeding $\tau = 3/2$. Our results establish a physical mechanism for scale-free dynamics beyond conventional criticality, providing a framework for understanding the non-universal avalanche statistics observed in earthquakes and dissipative systems.
\end{abstract}

\maketitle

{\it Introduction---}In many slowly driven systems, including earthquakes \cite{Gutenberg1956, Bak1989, Kawamura2012}, granular systems \cite{Lherminier2019, Houdoux2021, Daniels2008, Zadeh2019a}, sandpiles \cite{Frette1996, Ramos2009}, and  subcritical rupture \cite{Maloy2006, Xu2019, Baro2013, Stojanova2014, Bares2018}, energy accumulates gradually and is released through sudden events following power-law distributions. The consideration of a critical point as an attractor of the dynamics, introduced by the Self-Organized Criticality (SOC) \cite{BTW1987, Jensen1998} in 1987, established a direct link between scale-free avalanches and phase transitions \cite{Stanley1987}. Following this path, the mapping of avalanche dynamics onto a branching process in a Bethe lattice \cite{Alstrom1988} set the mean-field (MF) expectation of the size distribution of the events $P(s)\sim s^{-\tau}\exp{(-s/s_c)}$, with $\tau=3/2$ and $s_c$ scaling with system size. Scale-free avalanches spanning the system is the hallmark of  {\it criticality}, while dissipation shifts the cut-off $s_c$ to smaller values, destroying it \cite{Lauritsen1996}. Yet, the MF exponent $\tau=3/2$ remains remarkably robust \cite{Zapperi_1995, LeDoussal2009, Lauritsen1996, Dhar_1990}.

\begin{figure}[t!]
    \centering
    \includegraphics[width=9 cm]{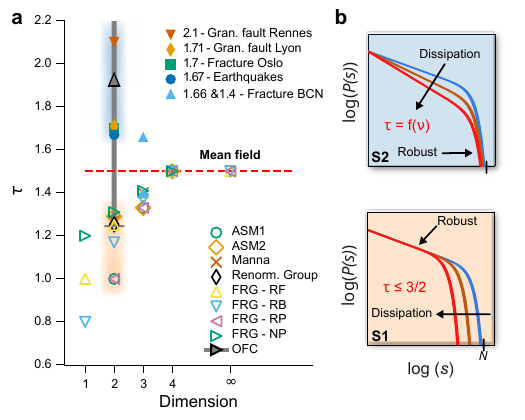}
        \vspace{-0.6cm}
    \caption{Two distinct behaviors. a) (open symbols) Theoretical predictions of the exponent values $\tau$ of the avalanche size distribution in Abelian sandpile models (ASM1 \cite{Zhang1989} and ASM2 \cite{Lubeck1997}), Manna model \cite{Chessa1999}, Renormalization Group  \cite{Vespignani1995} and Functional Renormalization Group (FRG) \cite{LeDoussal2009} with: Random field (RF), Random bond (RB), Random periodic (RP) and Non-periodic (NP). In UCD=4 and above, the MF solution of 3/2 is obtained. For two- and three-dimensions exponent values cluster below the value 3/2 in different universality classes. (solid symbols) OFC model and selection of experimental results in granular faults in Lyon \cite{Lherminier2019} and Rennes \cite{Houdoux2021} Labs, fracture in Oslo \cite{Maloy2006} and Barcelona (BCN) \cite{Xu2019}, and earthquake data \cite{Duplat2025, Navas2019} displaying exponent values above 3/2. b) Sketch of the two distinct behaviors of the avalanche size distributions when submitted to dissipation. {\bf S1}: robust $\tau\leqslant 3/2$ and cut-off $s_{max}$ that retreats with dissipation. {\bf S2}: robust cut-off $s_{max}$ proportional to the system size $N$ and $\tau$ values that increases with dissipation.}
    \label{fig:Lit}
        \vspace{-0.5cm}
\end{figure}

In Bethe lattices and MF approximations, the absence of spatial constraints renders the dynamics equivalent to infinite-dimensional systems. Below the upper critical dimension (UCD), loop formation during avalanche propagation complicates analytical treatment \cite{Dhar_1999}. Over two decades, numerous studies have sought to determine exponent values and classify avalanches into universality classes \cite{Lauritsen1996, Zapperi_1995, LeDoussal2009, Dhar_1990, Dhar_1999, Zhang1989, Lubeck1997, Chessa1999, Vespignani1995, Odor2004}. Zhang's 1989 relation \cite{Zhang1989}  $\tau = 2-2/d$, where $d$ is the system dimension, implied UCD=4, a result refined by subsequent numerical \cite{Lubeck1997, Chessa1999} and analytical \cite{Vespignani1995} work (Fig.~\ref{fig:Lit}a). Despite model-dependent discrepancies, consensus holds that in two and three dimensions $\tau<3/2$, as loops enhance local triggering and increase the likelihood of large avalanches. A major advance came in 2009 when Le Doussal and Wiese used the functional renormalization group to calculate exponents for various universality classes \cite{LeDoussal2009}. Their results are consistent with $\tau = 2-2/(d+\zeta)$, where $\zeta$ is the static roughness exponent of an elastic interface pinned by quenched disorder. Under random periodic disorder, with $\zeta = 0$, their results matched Zhang's prediction \cite{Zhang1989}.
This confirmed $\tau=3/2$ for $d\geqslant\text{UCD}$ and smaller values in lower dimensions (Fig.~\ref{fig:Lit}a). Although a clear consensus emerges from the literature, this is ---as  far
as we know--- the first report directly marking the limits of the theory. We refer to these applications of the critical-transitions formalism, which yield robust $\tau\leqslant 3/2$ across dimensions, as the {\bf S1} scenario (Fig.~\ref{fig:Lit}b).

\begin{figure*}[t!]
    \centering
    \includegraphics[width=\textwidth]{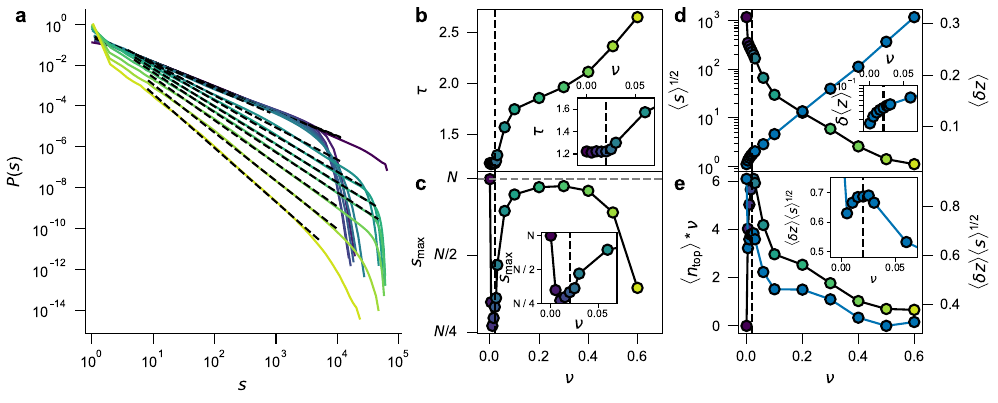}
       \vspace{-0.5cm}
    \caption{Size distributions \& trace dependence. a) Avalanche size distribution in the OFC model for different dissipation values ranging from 0 to 0.6 (dark to bright colors) for a system size L = 256. b) Power law exponents from the dashed lines in (a) as a function of dissipation. A vertical dashed line at $\nu = 2\%$ marks the threshold beyond which the exponent begins to increase. All insets provide a zoom into the low dissipation region. c) Maximum avalanche size as a function of dissipation. The horizontal dashed line is a visual indicator of the size of the system. d) Average linear size of the avalanches $\langle s\rangle^{1/2}$, scaling with the perimeter of the traces (color gradient) and mean value of the thickness $\delta z$ of the trace left by the avalanches (in blue). e) Average dissipated energy, $\langle n_\mathrm{top} \rangle \nu$, where $\langle n_\mathrm{top} \rangle$ is the mean number of topplings as a function of dissipation (indicated by a color gradient). The graph matches that of the product $\langle \delta z \rangle \langle s \rangle^{1/2}$, (in blue) indicating a direct relationship between dissipation and the traces left by the avalanches.}
    \label{fig:P(s)}
        \vspace{-0.5cm}
\end{figure*}

The branching process \cite{Alstrom1988}  captures the essence of this behavior: avalanches propagate stochastically between neighboring sites until growth ceases, with each event occurring independently, under statistically identical conditions. The absence of correlations between avalanches makes prediction impossible, which was a central claim of SOC. This notion, however, sparked intense debate when applied to earthquakes \cite{Geller1997, Wyss1997, Main1999}.
In the 1990s, limited seismic and experimental data were not precise enough to confront and guide theoretical efforts. Today, high-quality earthquake catalogs reveal exponents around $1.67$ \cite{Navas2019, Duplat2025}, while laboratory experiments consistently report $\tau>3/2$ across many decades of energy release \cite{Lherminier2019, Houdoux2021, Ramos2009, Maloy2006, Xu2019} (Fig.~\ref{fig:Lit}a and Appendix~\ref{AppxA}).  Moreover, observed exponent values vary with experimental conditions \cite{Lherminier2019, Houdoux2021, Ramos2009, Xu2019}, challenging the universality predicted by classical SOC theory.

This enduring discrepancy between theoretical predictions capped at $\tau=3/2$ and observations showing larger, non-universal exponents (Fig.~\ref{fig:Lit}a), highlights the need for a new unifying framework. Previous attempts to reconcile this gap, such as including relaxation mechanisms \cite{Aragon2012, Aragon2013} or fragmenting avalanches \cite{Jagla2013}, produce aftershock sequences that alter the global exponent. Yet, such dual distributions for aftershocks and background activity are absent in both experiments and natural seismicity \cite{Taroni2020}.

Here, we use the classical Olami–Feder–Christensen (OFC) earthquake model \cite{OFC1992, deArcangelis_pr-2016} as a minimal theoretical framework to demonstrate that scale-free avalanches can emerge through two fundamentally distinct mechanisms. We identify {\bf S1}, a genuinely critical regime characterized by robust avalanche-size exponents consistent with classical criticality, in which dissipation progressively reduces the distribution cutoff; and {\bf S2}, a quasi-critical regime characterized by non-universal, dissipation-dependent exponents larger than $3/2$, while the cutoff remains robust against variations in dissipation. The identification of these two distinct dynamical regimes, together with the crossover between them, provides a unified framework for understanding scale-free dynamics in driven dissipative systems.

{\it Revisiting the OFC model---}We performed extensive numerical simulations of the Olami--Feder--Christensen (OFC) model~\cite{OFC1992}, defined on a square lattice of $N=L\times L$ sites with open boundary conditions (Methods in~ \cite{SuppMat}). Each site $i$ is initialized with a random energy value $z_i \in[0,1)$, with a common threshold $z_{\mathrm{th}}\equiv1$. At every step, the system is loaded just enough for the site closest to the threshold to reach it. When a site topples, it resets to zero and transfers a fraction $\alpha$ of its value equally to its four nearest neighbors. The coupling parameter $\alpha\in[0,1/4]$ determines the dissipation, given by $\nu=1-4\alpha$. A toppling can trigger further topplings, and the process continues until all sites again satisfy $z_i<z_{\mathrm{th}}$. This propagation defines an avalanche, whose size $s$ is the total number of toppled sites, independent of how many times each site topples. 

{\it Different regimes in avalanche dynamics---}Figure~\ref{fig:P(s)}a shows the avalanche size distribution for a system of size $L=256$. In the conservative case ($\nu=0$) and at very low dissipation levels $\nu<2\%$, we obtain a robust exponent of  $\tau_c\simeq1.22$ (Fig.~\ref{fig:P(s)}b), while a cut-off, defined as the maximum avalanche size $s_{\text{max}}$, retreats with increasing dissipation (Fig.~\ref{fig:P(s)}c) consistent with the {\bf S1} scenario of Fig.~\ref{fig:Lit}b. Beyond $2\%$ dissipation, the exponent increases and the cut-off gradually approaches the size of the system, transitioning to the {\bf S2} scenario. Simulations at different system sizes (fig.~\ref{fig:Scaling})\cite{SuppMat} confirm the fact that $s_{\text{max}}\simeq N$ over the whole interval $\nu =0.1-0.4$. At higher dissipation values, the statistics is not sufficient in order to obtain clear cut-off values.

\begin{figure*}[t!]
    \centering
    \includegraphics[width=\textwidth]{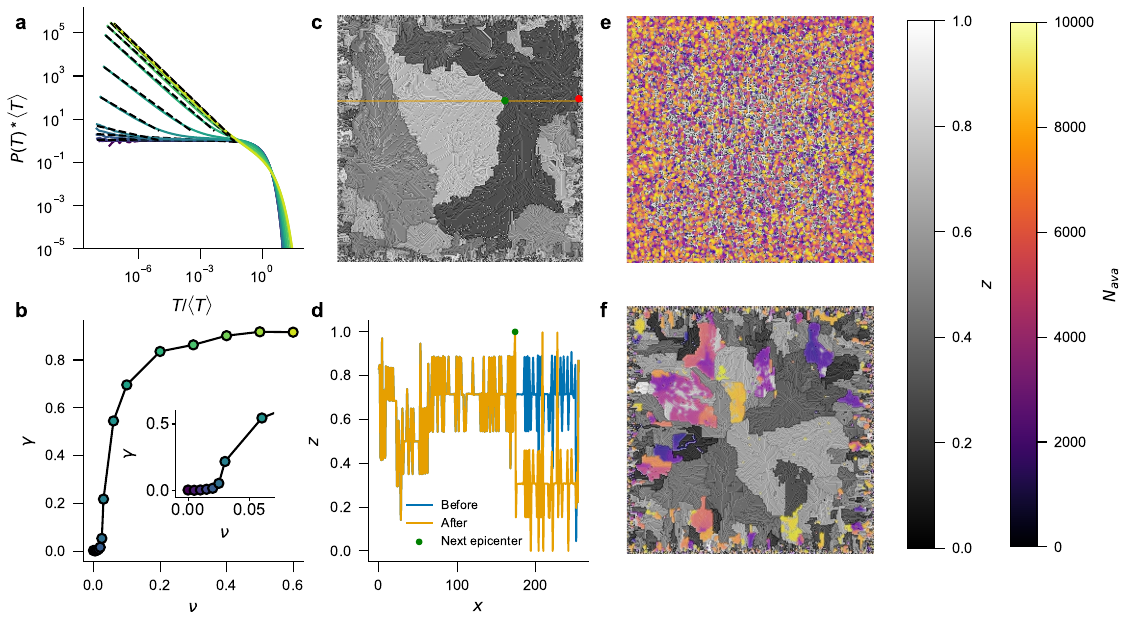}
    \caption{Memory effects \& patch formation.  a) Inter-event time distribution normalized by the mean intertime $\langle T \rangle$, for different dissipation values ranging from 0 to 0.6 (dark to bright colors), in a system of size L = 256. The black dashed line represent power law fits for each dissipation value, while the red dashed line correspond to a pure exponential decay. b) Exponent $\gamma$ of the power law fit from (a) as function of the dissipation. The inset provides a zoomed-in view of the low dissipation region. The vertical dashed line is a visual indicator for $2\%$ dissipation. Below $2\%$ distributions show a pure exponential decay, indicating a memoryless Poisson process, while above the transition, memory effects linked to temporal clustering of events, increased monotonically with dissipation. c) Spatial distribution of the energy after a large avalanche (darkest region). The grayscale represents the local energy levels.  Red and green points respectively represent the position of the previous and next epicenter. The orange line represent the position of the side-view shown in (d). d) Side view of the energy configuration, represented by the orange line in (c), before and after an avalanche. The green points represent the position of the next epicenter. e--f) Spatial distribution of the energy for $\nu$ = 0 (e) and 0.6 (f). Colored points represent a series of epicenters from dark to bright.}
    \label{fig:Intertime}
    \vspace{-0.5cm}
\end{figure*}

The possibility of system-spanning avalanches ($s_{\text{max}} \simeq N$) in dissipative systems, contrary to the predictions of conventional critical frameworks \cite{Lauritsen1996}, has remained a longstanding open question in the OFC model \cite{Bonachela2009, deArcangelis_pr-2016}. Here, based on more than 15 months of extensive simulations  \cite{SuppMat}, we demonstrate that $s_{\text{max}} \simeq N$ can indeed persist in the presence of dissipation, consistent with observations from earthquakes \cite{Navas2019, Duplat2025} and laboratory experiments \cite{Lherminier2019, DuplatOSLO2026}. 

We now turn to the physical mechanisms that allow system-spanning avalanches to persist despite dissipation.
Unlike branching processes on a Bethe lattice \cite{Alstrom1988}, OFC avalanches evolve over traces left by previous events, characterized by the \textit{trace thickness} $\delta z \equiv |\langle z_i \rangle_{i\in A}-\langle z_i \rangle_{i\notin A}|$, where $A$ denotes the sites involved in the preceding avalanche. Below $2\%$ dissipation, in the {\bf S1} regime, traces remain too thin to influence subsequent avalanches (inset, Fig.~\ref{fig:P(s)}d), resulting in wave-like, memoryless dynamics with Poissonian inter-event times \cite{corral2004}  (Fig.~\ref{fig:Intertime}a,b).

{\it Dynamical confinement creates the S2 regime---}Avalanche propagation leaves behind a trace surrounded by a rim whose length scales as $\langle s \rangle ^{1/2} $, and which frequently serves as a triggering region for subsequent events (Fig.~\ref{fig:Intertime}c and d)\cite{Kawamura2010}. As dissipation increases, $\langle \delta z \rangle$ rises sharply between 0 and 2\% dissipation (inset, Fig.~\ref{fig:P(s)}d). This increase outweighs —except in the conservative limit— the reduction in $\langle s \rangle ^{1/2}$, producing a maximum of $\langle \delta z \rangle \langle s \rangle ^{1/2}$ at 2\% dissipation (Fig.~\ref{fig:P(s)}e), which we identify as the transition between the {\bf S1} and {\bf S2} regimes. The quantity $\langle \delta z\rangle \langle s \rangle ^{1/2}$ characterizes an effective boundary layer of the avalanche traces, combining both their thickness and perimeter. Remarkably, it follows the same trend as the mean dissipated energy, $\langle n_{top} \rangle \nu$ (Fig.~\ref{fig:P(s)}e), indicating a direct relationship between the energy dissipated by an avalanche and the trace it leaves behind. Both quantities reach a maximum at the onset of memory effects \cite{corral2004,Lippiello2005, Kumar2020}.

In the conservative case ($\nu=0$), although 
$\langle \delta z \rangle \langle s \rangle ^{1/2}$
is largest, avalanche traces are broad but extremely shallow, not capable of pinning the dynamics. Accordingly, the triggering points (epicenters) remain spatially uncorrelated (Fig.~\ref{fig:Intertime}e), and avalanches explore the entire system. In contrast, in the {\bf S2} regime, increasing dissipation produces deeper traces that reshape the energy landscape in patches that evolve over extended periods. Patches are defined as cluster of sites with similar energy values. The activity is confined within spatial patches, where the dynamics become temporarily trapped (Fig.~\ref{fig:Intertime}f). This confinement allows the globally increasing energy to accumulate in other regions of the system, eventually synchronizing the energy values of neighboring patches needed for  triggering system-spanning events.

To disentangle the respective roles of dissipation and dynamical confinement in the {\bf S1}--{\bf S2} transition, we implemented a {\it democratic} algorithm in which the triggering site is selected uniformly at random and forced to topple, rather than choosing the most unstable site \cite{SuppMat}. This suppresses the preferential triggering at the rim of the previous avalanche trace and therefore prevents confinement within existing traces. The resulting dynamics retain {\bf S1} behavior over the entire dissipation range, with a robust exponent $\tau_c\simeq1.22$ and a cut-off that continues to decrease with dissipation, showing no indication of a transition (fig.~\ref{fig:Democratic}a)\cite{SuppMat}. Consistently, no spatial patches develop (fig.~\ref{fig:Democratic}b)\cite{SuppMat}. These results demonstrate that dissipation alone is insufficient to induce the transition. Instead, the presence of patches, which confine the dynamics within the traces of previous avalanches, is the essential ingredient driving the emergence of the {\bf S2} regime.

While dissipation destroys criticality in the classical {\bf S1} scenario (Fig.~\ref{fig:Lit}b), under the natural constraint that avalanches preferentially propagate over the traces left by previous events, it progressively deepens these traces until they become capable of confining subsequent avalanches. Beyond a threshold, this confinement gives rise to memory effects (Fig.~\ref{fig:Intertime}a), marking the transition to the {\bf S2} regime. Scale-free critical dynamics are recovered, defined here by the occurrence of system-spanning events ($s_{\text{max}}\simeq N$). 
However, since the avalanche-size distribution remains scale free and the mean avalanche size decreases with dissipation, the exponent $\tau$ necessarily increases, making system-spanning events progressively rarer. To understand how this scale-free organization is maintained despite dissipation, we first examine the critical properties of the system in the {\bf S2} regime.

\begin{figure*}[t!]
    \centering
    \includegraphics[width=13 cm]{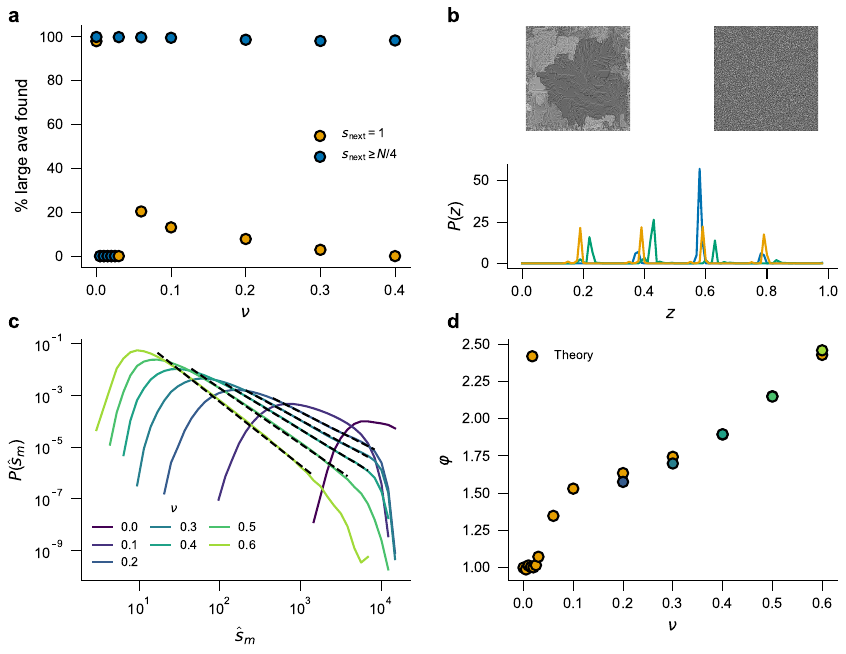}
    \caption{Transiting from critical to quasi-critical behavior.  a) Percentage of $s_c(\nu) \geq N/4$ for configurations immediately preceding large events ($s_{next}\geq N/4$), and those preceding very small events ($s_{next} = 1$) that are also far in time from any large event. For $\nu = 0$, large avalanches can be triggered from both types of configurations, indicating critical behavior. In contrast, for high dissipation values (\textbf{S2} scenario), the percentage of $s_{next} = 1$ configurations is about $20\%$ just after the transition and then decreases monotonically to zero as dissipation increases. b) For a given dissipation $\nu = 0.2$, the distribution of $z$-values --containing four peaks separated by $\alpha$-- is similar inside the patches (left panel) and in the case of one single domain (right panel), corresponding to a distribution of events with exponent $\tau_c \simeq 1.22$ in the democratic algorithm, suggesting a common intrinsic dynamics. b) Distributions of $s_c$ for different dissipation values ranging from 0 to 0.6 (dark to bright colors). They follow power laws for large dissipation values. d) Power-law exponents from the dashed lines in (c) as a function of dissipation (colors). Theoretical points correspond to $\varphi(\nu) = \tau(\nu) -\tau_c +1$, giving the expected behavior in the OFC model. }
    \label{fig:Transition}
\end{figure*}

{\it Permanent versus quasi-criticality---}The standard definition of criticality, $s_{\text{max}}\simeq N$, is a convenient shortcut to a more fundamental concept: in a genuinely critical system, a small avalanche must remain capable of cascading into a system-spanning event at essentially any time \cite{Alstrom1988, Main1999}. We refer to this property as the {\it criticality criterion} (CC), which captures the intrinsic fluctuation-driven nature of genuinely critical systems~\cite{Stanley1987, Carvalho2000, Bonachela_2009}.

By applying the \textit{maximum susceptibility} protocol (Appendix~\ref{Meth_Crit}), which measures the maximum possible avalanche size $s_c(\nu)$ under a  given configuration, we show that the OFC model in the {\bf S1} regime satisfies the CC.
We can consider that the system is therefore permanently critical.
In contrast, applying the same analysis to the {\bf S2} regime reveals that, before small events and far from large events --temporal zones that become increasingly frequent as the exponent $\tau$ increases-- the system is unable to generate a system-spanning avalanche (Fig.~\ref{fig:Transition}a). Therefore, the CC is not fulfilled. If we adopt the CC as a more stringent definition of criticality, then the {\bf S2} regime is not permanently critical. Instead, the system hovers around the critical point~\cite{Bonachela_2009}, briefly entering a critical state only before system-spanning events~\cite{Sammis2002, Ramos2010}. 

{\it A distribution of distances to criticality---}A simple theoretical description links the permanently critical {\bf S1} regime to the quasi-critical {\bf S2} regime. We will see how hovering around the critical point modifies the exponent of a power-law distribution, enabling a transition from the critical dynamics with $\tau=\tau_c$ to larger exponent values: consider a system exhibiting avalanches whose size distribution follows
\begin{equation}
\label{eq1}
    P(s)\sim s^{-\tau_c}e^{-s/s_c}
\end{equation}
\noindent where $\tau_c$ is the size exponent, and $s_c$ is the cut-off. The system is strictly critical only in the limit $s_c \rightarrow \infty$, which corresponds to $s_c \rightarrow N$ in the OFC model with dissipation $\nu =0$. Introducing dissipation in the {\bf S2} regime generates traces that confine the dynamics and temporally limit the extension of the largest possible avalanche.
The internal structure of each patch is similar to that of a memoryless system --obtained in the democratic algorithm-- but with the same level of dissipation (Fig.~\ref{fig:Transition}b).
Only small shifts in the $z$-distributions distinguish different patches. This suggests that inside each patch, Eq.~(\ref{eq1}) may describe the dynamics, rescaled down by the patch size. Furthermore, within each configuration or snapshot of the dynamics, Eq.~(\ref{eq1}) may also apply, with the cut-off  $s_c$ corresponding to the maximum avalanche size allowed in that specific configuration, which we can measure through $s_c(\nu)$.

Figure ~\ref{fig:Transition}c shows that for large dissipation values ($\nu\geqslant0.2$), the cut-off values are power-law distributed $P(s_c)\sim s_c^{-\varphi}$, with exponents dependent on dissipation $\varphi =\varphi(\nu)$.  We can then define a distance to the critical point (CP) as $d_{cp} \equiv 1/s_c$. This change in variables results in:
\begin{equation}
    \begin{aligned}
    P(d_{cp})\sim d_{cp}^{-\gamma},
    \\
     \quad \text{with~}  ~\gamma=2-\varphi<1
    \end{aligned}
     \label{eq2}
\end{equation}
\noindent Thus, the system hovers at a distance $d_{cp}$ from the CP, reflecting its dynamical trajectory through configurations characterized by different values of $s_c$.

The whole dynamics results from the integration over all different configurations (see details in \cite{SuppMat}):
\begin{equation}
    \begin{aligned}
        P(s) &= \int_0^\infty P(s|d_{cp})\,P(d_{cp})\,\mathrm{d}d_{cp} \sim s^{-\tau}, \\
        \quad \text{with}~&\tau(\nu) = \tau_c + 1 - \gamma(\nu)
    \end{aligned}
\label{eq3}
\end{equation}

\noindent which explains the transition from the critical dynamics with $\tau=\tau_c$ to a new distribution with exponent $\tau(\nu)$. A sketch of the model is shown in Fig.~\ref{fig:Conclusion}c.

As the system hovers around the CP, the dynamics pass through a series of configurations in which the maximum possible avalanche --measured as $s_c$-- does not scale with $N$ but is broadly distributed.

Combining Eqs. (\ref{eq2}) and (\ref{eq3}) we get
\begin{equation}
    \begin{aligned}
        \varphi(\nu)=\tau(\nu)-\tau_c+1
    \end{aligned}
\label{eq4}
\end{equation}
\noindent as the expected exponent values, which are in excellent agreement with the ones measured directly using $s_c(\nu)$ (Fig.~\ref{fig:Transition}d).

{\it Two routes to scale-free dynamics---}Using the OFC model as a minimal theoretical framework, we identify two distinct routes to scale-free avalanche dynamics. In the first, corresponding to the classical critical picture, the system is permanently critical, fluctuation-driven, and exhibit robust and universal exponents bounded by the mean field value $\tau=3/2$. 
In the second, dissipation and avalanche traces generate dynamical confinement. The system no longer remains permanently critical, but instead explores a broad distribution of distances to the critical point, producing quasi-critical scale-free dynamics with continuously varying exponents (Fig.~\ref{fig:Conclusion}).

This distinction resolves a longstanding paradox. Exponents larger than $3/2$ do not require abandoning scale-free organization or introducing separate aftershock physics; they can emerge naturally when sequential avalanches confine one another and the dynamics hover around criticality. The resulting framework links universal critical behavior and non-universal quasi-critical behavior within a single description.

Although demonstrated using the OFC model, the confinement mechanism relies only on the generic fact that sequential avalanches modify the conditions for subsequent ones. It is therefore a plausible general route to quasi-critical behavior in earthquakes and other slowly driven dissipative systems exhibiting crackling dynamics, as shown in experimental granular piles \cite{DuplatOSLO2026}. 

We thank Kay Wiesse, Andrei Fedorenko, Ezequiel Ferrero, Nicolas Bain and Olivier Cochet-Escartin for helpful discussions. This work was supported by the grant ANR-22-CE30-0046
\paragraph*{Data availability.}\url{https://figshare.com/s/f34daefd6005bfbc44ad}


\subsection*{End Matter}

\appendix

\section{Avalanche definition}\label{AppxA}

For elastic interfaces, where the exponent values have been analytically calculated with the tools of Functional Renormalization groups, avalanche size is conventionally defined as the total displacement integrated over the interface~\cite{LeDoussal2009},
\[
s=\int \Delta u(x)\mathrm{d}^{n}x\,
\]
where $n$ is the internal dimension of the interface and $\Delta u(\mathbf{x})$ is its local displacement during the event. This definition accounts for both the spatial extent of the avalanche and the displacement within it, providing a reference for comparing avalanche statistics across systems~\cite{Duplat2025}.

For earthquakes, the corresponding observable is the scalar seismic moment \cite{Kanamori1977},
\[
M_0=\mu\int_S D(x)\,\mathrm{d}S
   =\mu S\overline{D},
\]
where $\mu$ is the shear modulus, $S$ is the rupture area and $\overline{D}$ is the average relative slip across the fault. Seismic moment is therefore proportional to the total slip integrated over a two-dimensional fault surface. 

Radiated acoustic seismic energy can provide an equivalent measure of relative avalanche size when it scales linearly with seismic moment. For earthquakes, the empirical relation $E_s\simeq 1.6\times10^{-5}M_0$~\cite{Choy1995} provides an approximate conversion under this assumption.

\textit{Earthquake analogue experiments.}
 In laboratory experiments, acoustic-emission energy is likewise used to characterize avalanche size, including in the the fracture experiments in Barcelona~\cite{Xu2019}, and the granular fault experiments in Lyon~\cite{Lherminier2019} (Fig.~\ref{fig:Lit}a), where the linear scaling with mechanical energy was verified. Comparing the resulting exponents with moment-based exponents assumes that the measured acoustic energy is proportional to seismic moment over the range considered. In the granular fault experiment in Rennes~\cite{Houdoux2021} (Fig.~\ref{fig:Lit}a), avalanche size is quantified by an estimate of seismic moment. Finally, in the fracture experiment in Oslo~\cite{Maloy2006} (Fig.~\ref{fig:Lit}a), avalanche size corresponds to the area of clusters swept out by the advancing crack front. This area measures the displacement integrated along the one-dimensional front, consistently with the definition above.

\section{Critical and quasi-critical behaviors}\label{Meth_Crit}

\begin{figure*} [t!]
    \centering
    \includegraphics[width=\textwidth]{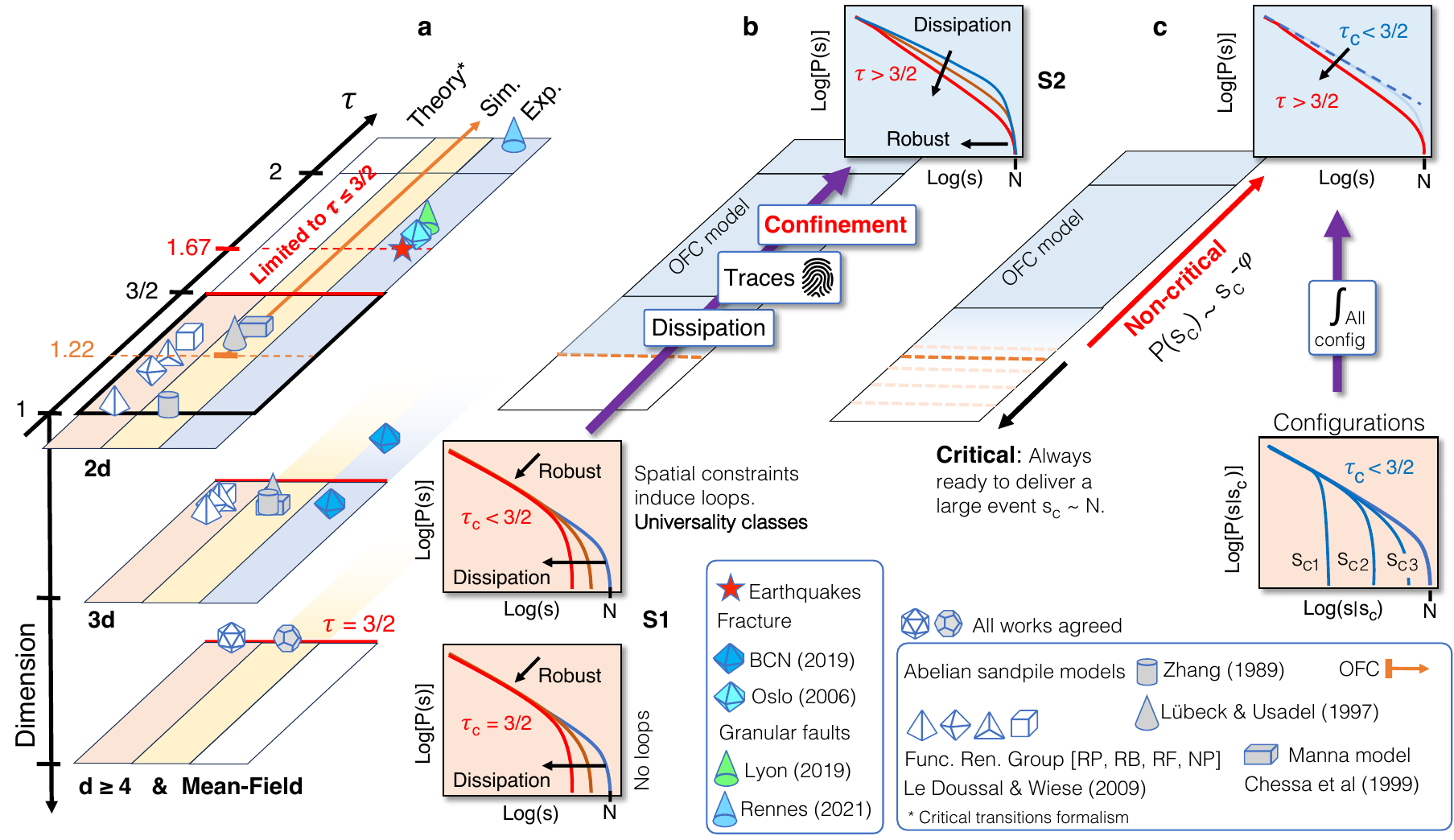}
    \caption{Summary. a) Compilation of theoretical, numerical, and experimental values of the avalanche-size exponent $\tau$ across spatial dimensions. The mean-field (MF) value $\tau=3/2$ applies only for $d \geqslant UCD=4$, and theoretical predictions are bounded by this $\tau=3/2$ value, whereas experimental measurements span a substantially broader range. References to the cited works are provided in the caption of Fig.~\ref{fig:Lit}. b) The OFC model exhibits two distinct dynamical regimes, {\bf S1} and {\bf S2}, and provides a mechanism for the transition between them. Increasing dissipation enlarges the traces (imprints) left by avalanches; the subsequent accumulation of these traces generates patches that confine epicenters and restore scale-invariant, system-spanning events, albeit with larger $\tau$ values that depend on the dissipation. c) In the conservative {\bf S1} regime, the system exhibits genuinely critical behavior ($s_c\sim N$). By contrast, for large dissipation values ({\bf S2} regime), the system is most of the time unable to generate large events, since the largest accessible avalanche sizes $s_c$ are themselves power-law distributed. The system is not critical; instead, it hovers around the critical point at a distance $d_{\mathrm{cp}}\equiv 1/s_c$. Integrating the avalanche-size distribution over all temporal configurations, characterized by different values of $s_c$, yields an effective distribution with a larger exponents, explaining the transition between {\bf S1} and {\bf S2} regimes.}
    \label{fig:Conclusion}
    \vspace{-0.5cm}
\end{figure*}

We define a protocol --labeled {\it Maximal susceptibility}-- to capture the maximum possible avalanche size $s_c (\nu)$ for a given configuration. To do this, during a classical OFC algorithm with dissipation $\nu$, and for a given configuration (snapshot), the system is first loaded following $\Delta z = z_{th} - z_\mathrm{max}$, where $z_\mathrm{max}$ is the value of the most unstable site. Then, each site is forced to topple, starting from the same given configuration, which --by following the standard redistribution process with dissipation $\nu$-- generates $N$ events of size $s_i$. The resulting maximum possible avalanche size is then defined as $s_c(\nu)=\max (s_i)$.

After verifying that forcing only 10\% of randomly selected sites that excludes the most unstable one, produces comparable results, this reduced protocol is adopted in place of toppling all sites. To evaluate the critical properties of the dynamics, we have introduced the criticality criterion (CC), according to which, statistically, at any moment, a small avalanche can cascade into a system-spanning event. For different dissipation values we have randomly selected 5000 configurations. Half of them just before a large ($s_{next}\geq N/4$) avalanche, and the other half preceding an event of size ($s_{next}=1$) and far from large events, i.e., no large event before a $\langle T \rangle /2$ time interval, and after a $\langle T \rangle /4$ time interval from the event. $\chi_{max}(\nu)$ is applied to all configurations. If $s_c(\nu)\geq N/4$, we consider that, in the analyzed configuration, any small avalanche can cascade into a system-spanning event. Fig.~\ref{fig:Transition}a shows that in the conservative case, from both initial configurations it is possible to generate large ($s\geq N/4$) avalanches, 100\% for $s_{next}\geq N/4$ configurations and about 98\% for $s_{next}=1$ ones. Statistically, the system is therefore almost always ready to generate a system-spanning event, fulfilling our CC. In contrast, in the {\bf S2} scenario, the probability of obtaining a large avalanche from a $s_{next}=1$ configuration is about 20\% just after the transition, and drops to zero as dissipation increases (Fig.~\ref{fig:Transition}a). This demonstrates that, for many configurations, a perturbation can no longer generate a system-spanning event. Our CC is not fulfilled.


\subsection*{Supplemental material}

\begin{figure*}[t!]
    \centering
    \includegraphics[width=0.7\textwidth]{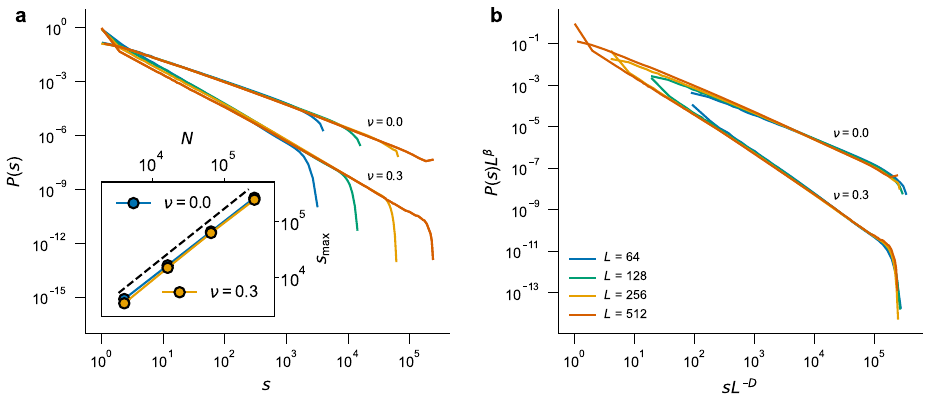}
    \caption{Size scaling in different regimes. a) Avalanche size distribution in the OFC, model for different system sizes ($L = 64, 128, 256, 512$) and for two different dissipation values $\nu = 0$ and $\nu = 0.3$. In both regimes the cut-off values scale with the size of the system (inset). b) Collapse of the distributions under finite-size scaling with $D=2.2$ and $\beta = 4.2$ and 2.8 for $\nu=0$ and $\nu = 0.3$ respectively.}
    \label{fig:Scaling}
\end{figure*}

\subsection*{Simulations in the OFC model}\label{Meth_sim}

We performed extensive numerical simulations of the Olami--Feder--Christensen (OFC) model~\cite{OFC1992}, defined on a square lattice of $N=L\times L$ sites with open boundary conditions. Each site $i$ is initialized with a random energy value $z_i\in[0,1)$ and a common threshold $z_{\mathrm{th}}\equiv1$. At each loading step, the system is uniformly driven by the minimal increment $\Delta z=z_{\mathrm{th}}-z_{\max}$, where $z_{\max}$ is the site closest to the threshold, ensuring that only one site becomes unstable at a time.

When a site reaches the threshold, it topples: its value is reset to zero and a fraction $\alpha$ of its energy is equally redistributed to its four nearest neighbors. The coupling parameter $\alpha\in[0,1/4]$ controls the dissipation in the system, quantified by $\nu=1-4\alpha$, with $\alpha=1/4$ corresponding to conservative bulk dynamics and $\alpha=0$ yielding trivial periodic behavior \cite{Ramos2006}. A toppling may trigger further instabilities, leading to a cascade that continues until all sites satisfy $z_i<z_{\mathrm{th}}$, defining an avalanche. To preserve the temporal structure of avalanches, a parallel updating scheme is employed \cite{Pinho_epjb-2000}. The avalanche size $s$ is defined as the number of distinct sites that topple, regardless of how many times each site topples.

Inter-event times $T$ are identified with the loading increments $\Delta z$, while avalanche durations are neglected. Memory effects are defined as deviations from a Poissonian inter-event time distribution \cite{corral2004}, presented in Fig.~3a,b. 

To ensure stationarity, an initial transient was discarded, ranging from $10^5$ avalanches for $\nu=0$ up to $2\times10^{11}$ avalanches for $\nu=0.6$. Depending on the dissipation, the total number of simulated avalanches varied between $10^8$ and $6 \times10^{12}$. The simulations required approximately 15 months of computation on a local server to obtain the statistics reported in Fig.~2 and Fig.~3. In particular, showing the possibility of system-spanning avalanches ($s_{\text{max}} \simeq N$) in dissipative systems (Fig.2c and Fig.~\ref{fig:Scaling}).

\begin{figure*}[t!]
    \centering
    \includegraphics[width=0.8\textwidth]{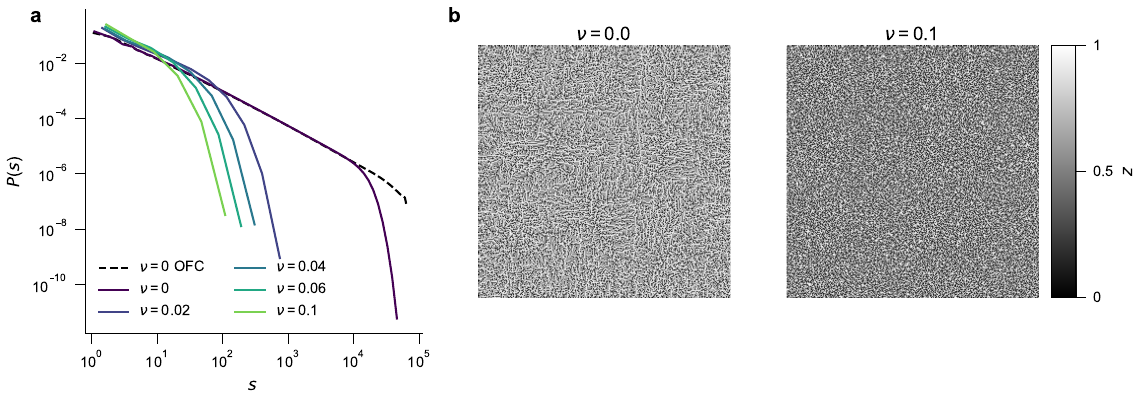}
    \caption{Democratic algorithm. a) Avalanche size distribution for different dissipation values. A \textbf{S1} behavior --with a robust $\tau = \tau_c \simeq 1.22$ and a cut-off that retreats with dissipation-- is obtained. The distribution for the conservative case in the non-modified OFC model (dashed line) is also plot for comparison. b) Snapshots of the structure under different dissipation values show that no patches appear.}
    \label{fig:Democratic}
\end{figure*}

\subsection*{Disentangling dissipation and dynamical confinement}\label{Meth_Dem}

By changing the rules of the OFC model it is possible to remove the correlations between events. We have run simulations using a democratic algorithm where the most unstable site is not the one that topples, but instead a random site is force to do it: its z value is set to 1, becoming unstable and toppling. Once the avalanche has finished, a new random sited is selected and forced to topple. This procedure suppresses the preferential triggering at the rim of the previous avalanche trace and therefore prevents confinement within existing traces. The resulting dynamics retain S1b behavior over the entire dissipation range, with a robust exponent $\tau_c \simeq 1.22$ and a cut-off that continues to decrease with dissipation, showing no indication of a transition (Fig.~\ref{fig:Democratic}a). Consistently, no spatial patches develop (Fig.~\ref{fig:Democratic}b).

\subsection*{Analytical model}\label{AppxB}

We consider a system in which the intrinsic avalanche dynamics follow a power-law distribution with exponent $\tau_c$, but where the maximum accessible avalanche size depends on the instantaneous configuration. Each configuration is characterized by a cutoff $s_c$, which quantifies its ability to sustain large avalanches. We define an effective distance to criticality as $d_{cp}\equiv 1/s_c$, such that configurations capable of generating large avalanches correspond to small values of $d_{cp}$.

For a given configuration, we assume that the avalanche-size distribution follows the intrinsic critical exponent $\tau_c$, with an exponential cutoff determined by $d_{cp}$:
\begin{equation}
    P(s|d_{cp}) =
    \frac{s^{-\tau_c}e^{-s d_{cp}}}
    {\mathcal{N}(d_{cp})},
    \label{eq7}
\end{equation}
where
$\mathcal{N}(d_{cp})=E_{\tau_c}(d_{cp})
=\int_1^\infty s^{-\tau_c}e^{-s d_{cp}}\,ds$
is the normalization constant for avalanche sizes $s\in[1,\infty)$, and $E_{\tau_c}$ denotes the generalized exponential integral.

In the quasi-critical regime, the system does not remain permanently near the critical point but explores configurations with different values of $d_{cp}$. Numerical simulations show that the distribution of accessible cutoff sizes follows $P(s_c)\sim s_c^{-\varphi}$, corresponding, through the change of variables $d_{cp}=1/s_c$, to $P(d_{cp})\sim d_{cp}^{-\gamma}$, with $\gamma=2-\varphi<1$.

The overall avalanche-size distribution is obtained by averaging the conditional distributions over all configurations:
\begin{subequations}
\begin{equation}
    P(s)=\int_0^\infty
    P(s|d_{cp})P(d_{cp})\,d d_{cp}.
    \label{eq8a}
\end{equation}
For large avalanche sizes, the integral is dominated by configurations with small $d_{cp}$, since the exponential factor suppresses contributions from configurations with $d_{cp}\gg 1/s$. Using $P(d_{cp})\sim d_{cp}^{-\gamma}$, we obtain, up to a multiplicative constant,
\begin{equation}
    P(s)\propto s^{-\tau_c}
    \int_0^\infty e^{-s d_{cp}}
    d_{cp}^{-\gamma}\,d d_{cp}.
    \label{eq8b}
\end{equation}
\end{subequations}
Here, we have used the fact that, for $\tau_c>1$, the normalization factor approaches a constant as $d_{cp}\rightarrow 0$:
$\mathcal{N}(d_{cp})\rightarrow 1/(\tau_c-1)$.
It therefore does not affect the asymptotic scaling exponent.

Introducing the change of variables $u=s d_{cp}$, Eq.~(\ref{eq8b}) becomes
\begin{equation}
    P(s)\propto
    s^{-(\tau_c+1-\gamma)}
    \int_0^\infty e^{-u}u^{-\gamma}\,du
    =\Gamma(1-\gamma)
    s^{-(\tau_c+1-\gamma)},
    \label{eq8c}
\end{equation}
where $\Gamma$ is the gamma function and the integral converges for $\gamma<1$.

We therefore obtain the asymptotic avalanche-size distribution
\begin{equation}
    P(s)\sim s^{-\tau},
    \qquad
    \tau=\tau_c+1-\gamma
    =\tau_c+\varphi-1>\tau_c.
    \label{eq9}
\end{equation}

This result shows that the non-universal exponent $\tau$ emerges from averaging intrinsic critical dynamics over a power-law distribution of accessible cutoff sizes. Thus, scale-free avalanche statistics can persist without the system remaining permanently critical.

\bibliography{Duplat_PRL}

@article{Maloy2006,
  	title = {Local Waiting Time Fluctuations along a Randomly Pinned Crack Front},
 	 author = {M\aa{}l\o{}y, Knut J\o{}rgen and Santucci, St\'ephane and Schmittbuhl, Jean and Toussaint, Renaud},
 	 journal = {Phys. Rev. Lett.},
	  volume = {96},
	  issue = {4},
	  pages = {045501},
 	 numpages = {4},
	  year = {2006},
	  month = {Jan},
	  publisher = {American Physical Society}
}

@ARTICLE{Baro2013,
  author = {Bar{\'o}, Jordi and Corral, {\'A}lvaro and Illa, Xavier and Planes,
	Antoni and Salje, Ekhard KH and Schranz, Wilfried and Soto-Parra,
	Daniel E and Vives, Eduard},
  title = {Statistical similarity between the compression of a porous material
	and earthquakes},
  journal = {Phys. Rev. Lett.},
  year = {2013},
  volume = {110},
  pages = {088702},
  number = {8},
  publisher = {APS}
}

@article{Stojanova2014,
 	 title = {High Frequency Monitoring Reveals Aftershocks in Subcritical Crack Growth},
 	 author = {Stojanova, M. and Santucci, S. and Vanel, L. and Ramos, O.},
 	 journal = {Phys. Rev. Lett.},
  	volume = {112},
	  pages = {115502},
 	 numpages = {5},
 	 year = {2014},
}

@ARTICLE{Gutenberg1956,
  author = {Gutenberg, Beno and Richter, Charles Francis},
  title = {Magnitude and energy of earthquakes},
  journal = {Ann. Geophys.},
  year = {1956},
  volume = {9},
  pages = {1--15},
  number = {1},
  publisher = {Istituto Nazionale di Geofisica e Vulcanologia}
}

@article {Choy1995,
author = {Choy, George L. and Boatwright, John L.},
title = {Global patterns of radiated seismic energy and apparent stress},
journal = { J. Geophys. Res. Solid Earth},
volume = {100},
number = {B9},
pages = {18205--18228},
year = {1995},
}

@Article{Bares2018,
	author={Bar{\'e}s, J.
	and Dubois, A.
	and Hattali, L.
	and Dalmas, D.
	and Bonamy, D.},
	title={Aftershock sequences and seismic-like organization of acoustic events produced by a single propagating crack},
	journal={Nature Communications},
	year={2018},
	volume={9},
	number={1},
	pages={1253},
}

@book{Stanley1987,
	  title = {Introduction to Phase Transitions and Critical Phenomena},
	  author = {Stanley, H. E.},
	   publisher = {Oxford Univ. Press},
	  address = {New York},
	  year = {1987},
}

@article{Bonachela_2009,
	year = 2009,
	month = {sep},
	publisher = {{IOP} Publishing},
	volume = {2009},
	number = {09},
	pages = {P09009},
	author = {Juan A Bonachela and Miguel A Mu{\~{n}}oz},
	title = {Self-organization without conservation: true or just apparent scale-invariance?},
	journal = {Journal of Statistical Mechanics: Theory and Experiment},
}

@article{Bak1989,
author = {Bak, Per and Tang, Chao},
title = {Earthquakes as a self-organized critical phenomenon},
journal = {Journal of Geophysical Research: Solid Earth},
volume = {94},
number = {B11},
pages = {15635-15637},
year = {1989}
}

@article{OFC1992,
  title = {Self-organized criticality in a continuous, nonconservative cellular automaton modeling earthquakes},
  author = {Olami, Zeev and Feder, Hans Jacob S. and Christensen, Kim},
  journal = {Phys. Rev. Lett.},
  volume = {68},
  pages = {1244--1247},
  year = {1992},
}

@article{Chessa1999,
  title = {Universality in sandpiles},
  author = {Chessa, Alessandro and Stanley, H. Eugene and Vespignani, Alessandro and Zapperi, Stefano},
  journal = {Phys. Rev. E},
  volume = {59},
  issue = {1},
  pages = {R12--R15},
  numpages = {0},
  year = {1999},
  month = {Jan},
  publisher = {American Physical Society}
}

@article{Lubeck1997,
  title = {Numerical determination of the avalanche exponents of the Bak-Tang-Wiesenfeld model},
  author = {L\"ubeck, S. and Usadel, K. D.},
  journal = {Phys. Rev. E},
  volume = {55},
  pages = {4095--4099},
  year = {1997},
}

@Article{Main1999,
author={Main, Ian},
title={Is the reliable prediction of individual earthquakes a realistic scientific goal?},
journal={Nature},
year={1999},
month={Feb},
day={25},
issn={1476-4687},
doi={10.1038/nature28107},
url={https://doi.org/10.1038/nature28107}
}

@article{Ramos2009,
  title = {Avalanche Prediction in a Self-Organized Pile of Beads},
  author = {Ramos, O. and Altshuler, E. and M\aa{}l\o{}y, K. J.},
  journal = {Phys. Rev. Lett.},
  volume = {102},
  pages = {078701},
  numpages = {4},
  year = {2009},
}

@article{Ramos2006,
  title = {Quasiperiodic Events in an Earthquake Model},
  author = {Ramos, O. and Altshuler, E. and M\aa{}l\o{}y, K. J.},
  journal = {Phys. Rev. Lett.},
  volume = {96},
  pages = {098501},
  numpages = {4},
  year = {2006},
 }

@article{Ramos2010,
title = {Criticality in earthquakes. Good or bad for prediction?},
journal = {Tectonophysics },
volume = {485},
number = {1-4},
pages = {321--326},
year = {2010},
author = {O. Ramos},
}

@article{Frette1996,
  author = {Frette, Vidar and Christensen, Kim and Malthe-S\o{}renssen, Anders and Feders, Jens and J\o{}ssang, Torstein and Meakin, Paul},
  title = {Avalanche dynamics in a pile of rice},
  journal = {Nature},
  year = {1996},
  volume = {379},
  pages = {49--52},
}

@article{BTW1987,
  title = {Self-organized criticality: An explanation of the 1/  \textit{f}  noise},
  author = {Bak, Per and Tang, Chao and Wiesenfeld, Kurt},
  journal = {Phys. Rev. Lett.},
  volume = {59},
  issue = {4},
  pages = {381--384},
  year = {1987},
   publisher = {American Physical Society}
}

@book{Jensen1998,
  author = {Jensen, H. J.},
  title = {Self-organized Criticality, Emergent Complex Behavior in Physical and Biological Systems},
  publisher = {Cambridge Univ. Press},
  address={Cambridge},
  year = {1998}
}

@article{Alstrom1988,
  title = {Mean-field exponents for self-organized critical phenomena},
  author = {Alstr\o{}m, Preben},
  journal = {Phys. Rev. A},
  volume = {38},
  issue = {9},
  pages = {4905--4906},
  numpages = {0},
  year = {1988}
}

@article{Geller1997,
author = {Geller, Robert J. and Jackson, David D. and Kagan, Yan Y. and Mulargia, Francesco}, 
title = {Earthquakes Cannot Be Predicted},
volume = {275}, 
number = {5306}, 
pages = {1616}, 
year = {1997}, 
journal = {Science},
}

@article{Wyss1997,
author = {Wyss, Max}, 
title = {Cannot Earthquakes Be Predicted?},
volume = {278}, 
number = {5337}, 
pages = {487--490}, 
year = {1997}, 
journal = {Science} 
}

@article{LeDoussal2009,
  title = {Size distributions of shocks and static avalanches from the functional renormalization group},
  author = {Le Doussal, Pierre and Wiese, Kay J\"org},
  journal = {Phys. Rev. E},
  volume = {79},
  issue = {5},
  pages = {051106},
  numpages = {34},
  year = {2009},
 }

@article{Lippiello2005,
  author={E. Lippiello and L. de Arcangelis and C. Godano},
  title={Memory in self-organized criticality},
  journal={Europhysics Letters},
  volume={72},
  number={4},
  pages={678},
  year={2005},
}

@article {Daniels2008,
author = {Daniels, Karen E. and Hayman, Nicholas W.},
title = {Force chains in seismogenic faults visualized with photoelastic granular shear experiments},
journal = {J. Geophys. Res. Solid Earth},
volume = {113},
number = {B11},
pages = {2156--2202},
year = {2008},
}

@article{Carvalho2000,
  title = {Self-Organized Criticality in the Olami-Feder-Christensen Model},
  author = {de Carvalho, Josu\'e X. and Prado, Carmen P. C.},
  journal = {Phys. Rev. Lett.},
  volume = {84},
  issue = {17},
  pages = {4006--4009},
  numpages = {0},
  year = {2000},
}

@article{Bonachela2009,
  author={Juan A Bonachela and Miguel A Muñoz},
  title={Self-organization without conservation: true or just apparent scale-invariance?},
  journal={Journal of Statistical Mechanics: Theory and Experiment},
  volume={2009},
  number={09},
  pages={P09009},
  year={2009},
}

@article{corral2004,
  title = {Long-Term Clustering, Scaling, and Universality in the Temporal Occurrence of Earthquakes},
  author = {Corral, \'Alvaro},
  journal = {Phys. Rev. Lett.},
  volume = {92},
  pages = {108501},
  numpages = {4},
  year = {2004},
}

@article{Lherminier2019,
  title = {Continuously Sheared Granular Matter Reproduces in Detail Seismicity Laws},
  author = {Lherminier, S. and Planet, R. and Levy dit Vehel, V.  and Simon, G. and Vanel, L. and M\aa{}l\o{}y, K. J. and Ramos, O.},
  journal = {Phys. Rev. Lett.},
  volume = {122},
  pages = {218501},
  numpages = {6},
  year = {2019},
}

@article{Zadeh2019a,
  title = {Crackling to periodic dynamics in granular media},
  author = {Abed Zadeh, Aghil and Bar\'es, Jonathan and Behringer, Robert P.},
  journal = {Phys. Rev. E},
  volume = {99},
  pages = {040901},
  numpages = {6},
  year = {2019},
}

@article{Kawamura2012,
  title = {Statistical physics of fracture, friction, and earthquakes},
  author = {Kawamura, Hikaru and Hatano, Takahiro and Kato, Naoyuki and Biswas, Soumyajyoti and Chakrabarti, Bikas K.},
  journal = {Rev. Mod. Phys.},
  volume = {84},
  pages = {839--884},
  numpages = {0},
  year = {2012},
}

@article{Odor2004,
  title = {Universality classes in nonequilibrium lattice systems},
  author = {\'Odor, G\'eza},
  journal = {Rev. Mod. Phys.},
  volume = {76},
  pages = {663-724},
  year = {2004},
}

@article{Zhang1989,
  title = {Scaling theory of self-organized criticality},
  author = {Zhang, Yi-Cheng},
  journal = {Phys. Rev. Lett.},
  volume = {63},
  issue = {5},
  pages = {470--473},
  numpages = {0},
  year = {1989},
  month = {Jul},
  publisher = {American Physical Society}
}

@article{Zapperi_1995,
  title = {Self-Organized Branching Processes: Mean-Field Theory for Avalanches},
  author = {Zapperi, Stefano and Lauritsen, Kent B\ae{}kgaard and Stanley, H. Eugene},
  journal = {Phys. Rev. Lett.},
  volume = {75},
  issue = {22},
  pages = {4071--4074},
  numpages = {0},
  year = {1995},
  month = {Nov},
  publisher = {American Physical Society}
}

@article{Lauritsen1996,
  title = {Self-organized branching processes: Avalanche models with dissipation},
  author = {B\ae{}kgaard Lauritsen, Kent and Zapperi, Stefano and Stanley, H. Eugene},
  journal = {Phys. Rev. E},
  volume = {54},
  issue = {3},
  pages = {2483--2488},
  numpages = {0},
  year = {1996},
  month = {Sep},
  publisher = {American Physical Society}
}

@article{Vespignani1995,
  title = {Renormalization approach to the self-organized critical behavior of sandpile models},
  author = {Vespignani, Alessandro and Zapperi, Stefano and Pietronero, Luciano},
  journal = {Phys. Rev. E},
  volume = {51},
  issue = {3},
  pages = {1711--1724},
  numpages = {0},
  year = {1995},
  month = {Mar},
  publisher = {American Physical Society}
}

@article{Dhar_1990,
year = {1990},
month = {oct},
publisher = {},
volume = {23},
number = {19},
pages = {4333},
author = {D Dhar and S N Majumdar},
title = {Abelian sandpile model on the Bethe lattice},
journal = {Journal of Physics A: Mathematical and General}
}

@article{Dhar_1999,
title = {The Abelian sandpile and related models},
journal = {Physica A: Statistical Mechanics and its Applications},
volume = {263},
number = {1},
pages = {4-25},
year = {1999},
note = {Proceedings of the 20th IUPAP International Conference on Statistical Physics},
issn = {0378-4371},
author = {Deepak Dhar}
}

@Article{Houdoux2021,
author={Houdoux, David
and Amon, Axelle
and Marsan, David
and Weiss, J{\'e}r{\^o}me
and Crassous, J{\'e}r{\^o}me},
title={Micro-slips in an experimental granular shear band replicate the spatiotemporal characteristics of natural earthquakes},
journal={Communications Earth {\&} Environment},
year={2021},
month={May},
day={14},
volume={2},
number={1},
pages={90},
issn={2662-4435}
}

@article{Pinho_epjb-2000,
	title = {Sequential updates for non-abelian {SOC} models},
	volume = {18},
	issn = {1434-6028},
	pages = {479--484},
	number = {3},
	journal = {The European Physical Journal B},
	shortjournal = {Eur. Phys. J. B},
	author = {Pinho, S.T.R. and Prado, C.P.C.},
	urldate = {2023-10-11},
	date = {2000-12},
    year = {2000},
	langid = {english},
}

@article{Kanamori1977,
author = {Kanamori, Hiroo},
title = {The energy release in great earthquakes},
journal = {J. Geophys. Res.},
volume = {82},
number = {20},
pages = {2981--2987},
year = {1977}
}

@article{deArcangelis_pr-2016,
	title = {Statistical physics approach to earthquake occurrence and forecasting},
	volume = {628},
	issn = {03701573},
	
	pages = {1--91},
	journal = {Physics Reports},
	shortjournal = {Physics Reports},
	author = {De Arcangelis, Lucilla and Godano, Cataldo and Grasso, Jean Robert and Lippiello, Eugenio},
	urldate = {2024-05-13},
	date = {2016-04},
    year = {2016},
	langid = {english},
}

@article{Xu2019,
  title = {Criticality in failure under compression: Acoustic emission study of coal and charcoal with different microstructures},
  author = {Xu, Yangyang and Borrego, Angeles G. and Planes, Antoni and Ding, Xiangdong and Vives, Eduard},
  journal = {Phys. Rev. E},
  volume = {99},
  issue = {3},
  pages = {033001},
  numpages = {12},
  year = {2019},
  month = {Mar},
  publisher = {American Physical Society}
}

@article{Navas2019,
  title = {Universality of power-law exponents by means of maximum-likelihood estimation},
  author = {Navas-Portella, V\'{\i}ctor and Gonz\'alez, \'Alvaro and Serra, Isabel and Vives, Eduard and Corral, \'Alvaro},
  journal = {Phys. Rev. E},
  volume = {100},
  issue = {6},
  pages = {062106},
  numpages = {14},
  year = {2019},
  month = {Dec},
  publisher = {American Physical Society}
}

@article{Aragon2012,
  title = {Seismic cycles, size of the largest events, and the avalanche size distribution in a model of seismicity},
  author = {Arag\'on, L. E. and Jagla, E. A. and Rosso, A.},
  journal = {Phys. Rev. E},
  volume = {85},
  issue = {4},
  pages = {046112},
  numpages = {10},
  year = {2012},
  month = {Apr},
  publisher = {American Physical Society}
}

@article{Aragon2013,
    author = {Arag\'on, L. E. and Jagla, E. A.},
    title = {Spatial and temporal forecasting of large earthquakes in a spring-block model of a fault },
    journal = {Geophysical Journal International},
    volume = {195},
    number = {3},
    pages = {1763-1772},
    year = {2013},
    month = {09},
    issn = {0956-540X}
}

@article{Jagla2013,
  title = {Forest-Fire Analogy to Explain the $b$ Value of the Gutenberg-Richter Law for Earthquakes},
  author = {Jagla, E. A.},
  journal = {Phys. Rev. Lett.},
  volume = {111},
  issue = {23},
  pages = {238501},
  numpages = {5},
  year = {2013},
  month = {Dec},
  publisher = {American Physical Society}
}

@article{Taroni2020,
    author = {Taroni, Matteo and Akinci, Aybige},
    title = {Good practices in PSHA: declustering, b-value estimation, foreshocks and aftershocks inclusion; a case study in Italy},
    journal = {Geophysical Journal International},
    volume = {224},
    number = {2},
    pages = {1174-1187},
    year = {2020},
    month = {09},
    issn = {0956-540X}
}

@Article{Kumar2020,
author={Kumar, Pinaki
and Korkolis, Evangelos
and Benzi, Roberto
and Denisov, Dmitry
and Niemeijer, Andr{\'e}
and Schall, Peter
and Toschi, Federico
and Trampert, Jeannot},
title={On interevent time distributions of avalanche dynamics},
journal={Scientific Reports},
year={2020},
month={Jan},
day={17},
volume={10},
number={1},
pages={626},
issn={2045-2322}
}

@article{Kawamura2010,
  title = {Asperity characteristics of the Olami-Feder-Christensen model of earthquakes},
  author = {Kawamura, Hikaru and Yamamoto, Takumi and Kotani, Takeshi and Yoshino, Hajime},
  journal = {Phys. Rev. E},
  volume = {81},
  issue = {3},
  pages = {031119},
  numpages = {10},
  year = {2010},
  month = {Mar},
  publisher = {American Physical Society}
}

@article{Sammis2002,
author = {C. G. Sammis  and D. Sornette },
title = {Positive feedback, memory, and the predictability of earthquakes},
journal = {Proceedings of the National Academy of Sciences},
volume = {99},
number = {suppl\_1},
pages = {2501-2508},
year = {2002}
}

@misc{DuplatOSLO2026,
      title={OFC-like Behavior in Experimental Granular Piles}, 
      author={K. Duplat and A. Douin and E. Saurety and G. Simon and E. Altshuler and K.J. M\aa{}l\o{}y and O. Cochet-Escartin and  O. Ramos},
      year={2026},
      note={arXiv:2609.14166 [cond-mat.stat-mech]}
}

@misc{Duplat2025,
      title={Gutenberg-Richter-like relations in physical systems}, 
      author={K. Duplat and G. Varas and O. Ramos},
      year={2025},
      note={arXiv:2512.17615 [cond-mat.stat-mech]}
}

@misc{SuppMat,
      note={Supplemental Material}
}

\end{document}